\documentclass[aps,prl,superscriptaddress,amsmath,groupedaddress,twocolumn]{revtex4}

\usepackage{color,hangcaption,hhline,psfrag,rotating,amssymb}
\usepackage[hang,nooneline]{subfigure}
\usepackage{dcolumn}

\begin{document}
\title{Precise charm to strange mass ratio and light quark masses from full lattice QCD}

\author{C. T. H. Davies}
\email[]{c.davies@physics.gla.ac.uk}
\affiliation{Department of Physics and Astronomy, University of Glasgow, Glasgow, G12 8QQ, UK}
\author{C. McNeile}
\affiliation{Department of Physics and Astronomy, University of Glasgow, Glasgow, G12 8QQ, UK}
\author{K. Y. Wong}
\affiliation{Department of Physics and Astronomy, University of Glasgow, Glasgow, G12 8QQ, UK}
\author{E. Follana}
\affiliation{Departamento de F\'{\i}sica Te\'{o}rica, Universidad de Zaragoza, E-50009 Zaragoza, Spain}
\author{R. Horgan}
\affiliation{DAMTP, Cambridge University, Wilberforce Road, Cambridge, CB3 0WA, UK}
\author{K. Hornbostel}
\affiliation{Southern Methodist University, Dallas, Texas 75275, USA}
\author{G. P. Lepage}
\affiliation{Laboratory of Elementary-Particle Physics, Cornell University, Ithaca, New York 14853, USA}
\author{J. Shigemitsu}
\affiliation{Department of Physics, The Ohio State University, Columbus, Ohio 43210, USA}
\author{H. Trottier}
\affiliation{Physics Department, Simon Fraser University, Vancouver, BC, Canada}

\collaboration{HPQCD collaboration}
\homepage{http://www.physics.gla.ac.uk/HPQCD}
\noaffiliation

\date{\today}

\begin{abstract}
By using a single formalism to handle charm, strange
and light valence quarks in full lattice QCD for the first time, we are able to determine ratios of 
quark masses to 1\%. For $m_c/m_s$ we obtain 
11.85(16), an order of magnitude more precise than the current PDG average. 
Combined with 1\% determinations of the charm quark mass now possible this 
gives $\overline{m}_s(2{\rm GeV}) =$ 92.4(1.5) MeV. The MILC result for $m_s/m_l = 27.2(3) $ yields 
$\overline{m}_l(2{\rm GeV})$ = 3.40(7) MeV for the average of $u$ and $d$ quark masses.  
\end{abstract}


\maketitle

{\it Introduction}. -- The masses of $u$, $d$ and $s$ quarks are some of the least well-known parameters 
of the Standard Model. 
Even the most inaccurate lepton mass (that of the $\tau$) is known 
to better than 0.01\% and yet errors on light quark masses of 30\%
are quoted in the Particle Data Tables~\cite{pdg09}. 
The reason for the mismatch is the confinement property
of the strong force that obscures the connection between 
the properties of the quark constituents and the hadron physics that is 
accessible to experiment. To make this connection requires 
accurate calculations in QCD and accurate experimental results 
for appropriate hadronic quantities. A method particularly 
well-suited to this is lattice QCD. 
Here we will demonstrate its use by
determining $m_c/m_s$ to 1\% and obtaining as a result 1.5\% errors
for light quark masses, which 
brings them almost into line with those of 
heavy quarks. 

Heavy quark masses, $m_Q$, can be determined accurately because $\alpha_s(m_Q)$ 
is relatively small. 
1\% 
errors for 
charm and bottom quark masses have recently become possible 
using $\cal{O}$$(\alpha_s^3)$ calculations in 
 QCD perturbation theory for the heavy quark vacuum `bubble'~\cite{pert}
and therefore for  
the energy-derivative (or time) 
moments of correlation functions for
a heavy quark-antiquark pair at zero momentum.
Since the scale of $\alpha_s$ is naturally related to the relevant heavy 
quark mass, the expressions can be evaluated accurately. 
To extract the quark mass the perturbative result is compared to
a nonperturbative determination containing information from experiment. 
For a $1^{--}$ $Q\overline{Q}$ configuration 
moments of the 
experimentally measured cross-section for ($e^+e^- \rightarrow \gamma^* \rightarrow {\mathrm{hadrons}}$) 
can be used after isolating the heavy quark contribution 
and using dispersion relations~\cite{kuhn09}. 
Alternatively the time-moments for heavy quark 
current-current correlation functions of various $J^{PC}$ 
can be directly determined 
in lattice QCD calculations that have been tuned 
so that a charmonium or bottomonium mass agrees with experiment~\cite{mcjj,mylat08}. 
The time moments must be extrapolated to the 
zero lattice spacing (continuum) limit before the comparison to QCD 
perturbation theory. 
These two methods give results that agree, with 
1\% errors for $m_c(3 {\mathrm{GeV}})$ in the $\overline{MS}$ scheme. 
The more traditional `direct' lattice QCD method, although 
somewhat less accurate, also gives results in good agreement~\cite{mchighbeta}. 
We can conclude from this that $m_c$ is now accurately known. 

The strange quark mass, $m_s$, being much smaller, cannot be determined this way 
and is 
poorly known at present. 
Instead of a direct determination of $m_s$, however, 
we can use the leverage of an accurate result for the 
ratio $m_c/m_s$ combined with the accurate 
$m_c$ above~\cite{gasserleutwyler}. 
But simple ratios of hadron mass differences give unreliable estimates 
of $m_c/m_s$. Two such estimates:
\begin{equation}
\frac{m(B_c)-m(B_u)}{m(B_s)-m(B_u)} = 11;
\frac{m(\Sigma_c)-m(N)}{m(\Sigma)-m(N)} = 6
\end{equation}
differ by almost a factor of 2.
The ratio of $m_s/m_l$ (where $m_l = (m_u+m_d)/2$) 
is known to about 10\% from ratios of squared masses of $K$ and $\pi$ mesons
using SU(3) chiral perturbation theory~\cite{pdg09}. 
Clearly neither ratio is determined well enough this way to provide the 
accuracy we need, because 
the relationship between hadron mass and well-defined running quark mass 
is more complicated than these simple ratios must assume. 

Lattice QCD, on the other hand, can give very accurate 
results for the ratio of two quark masses but only if the same formalism 
is used for both quarks. This has already been used 
to give accurate results for $m_s/m_l$, although neither $m_s$ nor 
$m_l$ is very well determined.
Here, for the first time, we give an accurate result for $m_c/m_s$ 
by using the same formalism for charm, strange and light quarks 
and this enables us to 
cascade the accuracy of the heavy quark mass down 
to the light quarks. 

{\it The Lattice QCD calculation.} -- Lattice QCD gives direct access to quark 
masses through
the lattice QCD Lagrangian. 
Tuning of the masses is done by calculating 
an appropriate hadron mass and adjusting the 
quark mass until the hadron mass agrees with experiment.  
Experimental measurements
of appropriate hadron masses 
are extremely accurate in most cases, 
with errors at the level of tenths or hundredths of a percent. To make 
maximum use of this precision we need 
to calculate the hadron mass in lattice QCD with small 
statistical and systematic errors. 
In particular it requires the 
full effect of sea quarks in the hadron to be included. 
This is now possible in lattice QCD~\cite{ratio}. Fixing the 
four quark masses ($m_l$, $m_s$, $m_c$, $m_b$) from 
four `gold-plated' hadrons ($\pi$, $K$, $\eta_c$, $\Upsilon$) enables 
other quantities to be calculated with errors of a few 
percent and agreement with experiment is obtained~\cite{ratio, lp07}. 
This is an important test that QCD, 
with only one scale parameter and one mass parameter per quark 
flavor,
describes the full range of hadron physics consistently. 

The lattice quark mass is a perfectly well-defined
running quark mass. However, it is scheme-dependent and so 
varies with the discretisation of the Dirac equation used 
in the lattice calculation. 
For wider applicability it is more useful to convert the lattice
quark mass to a standard continuum scheme such as $\overline{MS}$.  
This renormalization has been a major source of systematic error in previous 
determinations of light and strange quark masses. 
The best existing result for $m_s(2 {\mathrm{GeV}})$, with a 7\% error, uses the 
direct method of converting the tuned quark 
mass in the lattice QCD Lagrangian to the $\overline{MS}$ scheme 
using $\alpha_s^2$ lattice QCD perturbation theory~\cite{ourstrange}. 
The error is dominated by the error in the renormalization 
and it is the error
that we will remove here, by instead determining 
$m_c/m_s$ accurately.  
The Highly Improved Staggered Quark 
action~\cite{hisq,fds} allows us to use the same 
discretization of QCD for both charm and strange 
quarks 
because it is a fully relativistic `light quark' action 
that can also be used for charm quarks. 
Then the mass renormalisation factor cancels 
in the quark mass ratio. 

We work with eight different ensembles of gluon field configurations provided by 
the MILC collaboration. These include the effect 
of $u$, $d$ and $s$ sea quarks using the improved 
staggered quark (asqtad) formalism using the fourth root `trick'. 
This procedure, although `ugly', appears to be a valid 
discretization of QCD~\cite{sharpe,milcreview,kronfeld,golterman}. 
Tests include studies of the Dirac operator and comparisons 
to effective field theories.  
Configurations are available with 
large spatial volumes ($> 2.4({\rm fm})^3$) at multiple 
values of the light sea masses (using $m_u = m_d = m_l$) 
and for a wide range of 
values of the lattice spacing, $a$. 
We use configurations at five values of $a$
between 0.15 fm and 0.05 fm with parameters as listed 
in Table~\ref{tab:params}. 

\begin{table}
\begin{tabular}{lllllllll}
\hline
\hline
Set & $\beta$ & $r_1/a$ & $au_0m_{0l}^{asq}$ & $au_0m_{0s}^{asq}$ & $L/a$ & $T/a$ & $N_{conf}\times N_{t}$ \\
\hline
1 & 6.572 & 2.152(5) & 0.0097 & 0.0484 & 16 & 48 & $631 \times 2$\\
2 & 6.586 & 2.138(4) & 0.0194 & 0.0484 & 16 & 48 & $631 \times 2$\\
\hline
3 & 6.76 & 2.647(3) & 0.005 & 0.05 & 24 & 64 & $678 \times 2$ \\
4 & 6.76 & 2.618(3) & 0.01 & 0.05 & 20 & 64 & $595 \times 2 $ \\
\hline
5 & 7.09 & 3.699(3) & 0.0062 & 0.031 & 28 & 96 & $566 \times 4$ \\
6 & 7.11 & 3.712(4) & 0.0124 & 0.031 & 28 & 96 & $265 \times 4$ \\
\hline 
7 & 7.46 & 5.296(7) & 0.0036 & 0.018 & 48 & 144 & $201 \times 2$ \\
\hline
8 & 7.81 & 7.115(20) & 0.0028 & 0.014 & 64 & 192 & $208 \times 2 $ \\
\hline
\hline
\end{tabular}
\caption{Ensembles (sets) of MILC configurations used, with gauge coupling $\beta$, 
size $L^3 \times T$ and sea 
masses ($\times$ tadpole parameter $u_0$) 
$m_{0l}^{asq}$ and $m_{0s}^{asq}$. 
The lattice spacing values in units of $r_1$ after `smoothing'
are given in column 3~\cite{milcreview}. 
Column 8 
gives the number of configurations and time sources per configuration 
used for calculating correlators. }
\label{tab:params}
\end{table}

\begin{table}
\begin{tabular}{llllll}
\hline
\hline
Set & $am_{0c}$ &   $1+\epsilon$ & $am_{\eta_c}$ & $am_{0s}$ & $am_{\eta_s}$ \\
\hline
1 & 0.81 & 0.665 & 2.19381(16) & 0.061 & 0.50490(36) \\
 & 0.825 &  0.656 & 2.22013(15) & 0.066 & 0.52524(36) \\ 
 & 0.85 &  0.641 & 2.26352(15) & 0.080 & 0.57828(34) \\
2 & 0.825 &  0.656 & 2.21954(13) & 0.066 & 0.52458(35) \\
\hline
3 & 0.65 &  0.762 & 1.84578(8) & 0.0537 & 0.43118(18) \\
4 & 0.63 & 0.774 & 1.80849(11) & 0.0492 & 0.41436(23) \\
 & 0.66 &  0.756 & 1.86674(19) & 0.0546 & 0.43654(24) \\
 & 0.72 &  0.72 & 1.98114(15) & 0.05465 & 0.43675(24) \\
 & 0.753 &  0.70 & 2.04293(10) & 0.06 & 0.45787(23) \\
 &  &   & & 0.063 & 0.46937(24) \\
\hline
5 & 0.413 & 0.893 & 1.28057(7) & 0.0337 & 0.29413(12) \\
  & 0.43 &  0.885 & 1.31691(7) & 0.0358 & 0.30332(12) \\
  & 0.44 &  0.88 & 1.33816(7) & 0.0366 & 0.30675(12) \\  
  & 0.45 &  0.875 & 1.35934(7) & 0.0382 & 0.31362(14) \\
6 & 0.427 &  0.885 & 1.30731(10) & 0.03635 & 0.30513(20) \\
\hline
7 & 0.273 & 0.951  & 0.89932(12) & 0.0228 & 0.20621(19) \\
  & 0.28 & 0.949 & 0.91551(9) & 0.024 & 0.21196(13) \\
\hline
8 & 0.195 & 0.975 & 0.67119(6) & 0.0165 & 0.15484(14) \\
  &  &  &  & 0.018 & 0.16209(17) \\
\hline
\hline
\end{tabular}
\caption{ Results for the masses in lattice units of the goldstone pseudoscalars made 
from valence HISQ charm or strange quarks on the different 
MILC ensembles enumerated in Table~\ref{tab:params}. Columns 2 and 3 give 
the corresponding bare charm quark mass, 
and Naik coefficient respectively. Column 6 gives the bare strange 
quark mass ($\epsilon=0$ in that case). 
 }
\label{tab:charmmass}
\end{table}

On these configurations we have calculated quark propagators 
for charm quarks, strange quarks and light quarks (again
$m_u=m_d=m_l$) using the HISQ action. The numerical speed of 
HISQ means that we have been able to use several
nearby quark masses for charm and strange to allow accurate
interpolation to the correct values. Table~\ref{tab:charmmass} 
gives masses for the goldstone pseudoscalar mesons made from 
either a charm quark-antiquark pair or a strange one (the $\eta_c$ 
and the $\eta_s$), which are used for tuning. 
In the charm case, as well as the quark mass, we list 
the coefficient of the
`Naik' term in 
the HISQ action that corrects for discretisation errors through 
$(am_{0c})^4$.
The quark propagators are generated from random wall sources
and the goldstone mesons have good signal/noise properties 
so the meson masses can be determined to high precision 
using a standard multi-exponential fit~\cite{bayesfits}. 

\begin{figure}[h]
\includegraphics[width=8.5cm]{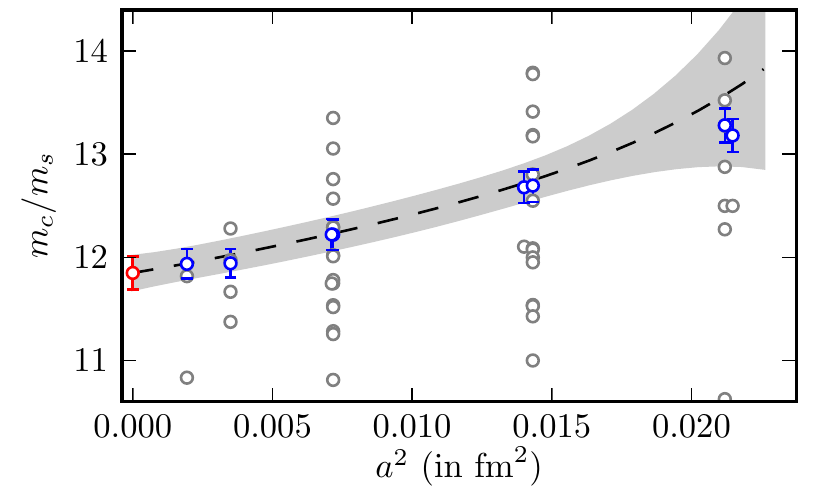}
\caption{ Grey points show the raw data for every ratio of $m_c/m_s$ on each 
ensemble (Table~\ref{tab:charmmass}); these ratios are fit to eq.~\ref{eq:fit}. 
The dashed line and associated grey error band (and red point at $a=0$) show our 
extrapolation of the resulting tuned $m_c/m_s$ to the continuum limit. 
Blue points with error bars are from a simple interpolation, separately 
for each ensemble, to the correct $m_c/m_s$, and are shown for illustration.}
\label{fig:mcmsplot}
\end{figure}

The meson masses can be converted to physical units 
with a determination of the lattice spacing. 
On an ensemble by ensemble basis this is taken from 
a parameter in the heavy quark potential called $r_1$.
Values for $r_1/a$ 
determined by the MILC collaboration~\cite{milcreview} are given in 
Table~\ref{tab:params}. They have errors of 0.3-0.5\%. 
The physical value for $r_1$ must then be obtained by comparing 
to experimentally known quantities and we use the 
value 0.3133(23) fm obtained from a set of four such quantities, 
tested for consistency in the continuum limit~\cite{newr1, massnote}. 

Using the information about meson masses that we have on each ensemble
we can interpolate to the correct ratio for 
$am_{0c}$ and $am_{0s}$ using appropriate continuum values 
for the masses of the $\eta_c$ and $\eta_s$. 
We correct the experimental value of $m_{\eta_c}$ of 2.9803 GeV
to 
$m_{\eta_c, \mathrm{phys}}$ = 2.9852(34) GeV. 
This allows for electromagnetic effects (2.4 MeV)~\cite{newr1} and 
$\eta_c$ annihilation to gluons (2.5MeV)~\cite{hisq}, both 
of which are missing from our calculation, so increasing the 
$\eta_c$ mass. We take a 50\% error on each 
of these corrections and also increase the experimental error to 
3 MeV to allow for the spread of results from different 
$\eta_c$ production mechanisms~\cite{pdg09}. 
Since 
the total shift is only around 0.2\% of the $\eta_c$ mass it has a
negligible effect as can be seen from our error budget below. 

The $\eta_s$ is not a physical particle in the real world because of 
mixing with other flavor neutral combinations to make the $\eta$ and $\eta^{\prime}$. 
However, in lattice QCD, the particle calculated (as here) from 
only `connected' quark propagtors does not mix and is a well-defined 
meson. Its mass must be determined by relating its properties to 
those of mesons such as the $\pi$ and $K$ that do appear in experiment. 
From an analysis of the lattice spacing and $m_l$-dependence of the 
$\pi$, $K$, and $\eta_s$ masses we conclude that the 
value of the $\eta_s$ mass in the continuum and physical $m_l$ limits 
is 0.6858(40) GeV~\cite{newr1}.  

The connection between the $\overline{MS}$ mass at a scale 
$\mu$ and the lattice bare quark mass is given by~\cite{mctree, ourstrange}:
\begin{eqnarray}
&&\overline{m}(\mu) = \frac{am_0}{a} Z_m(\mu a, m_0a), \\ \nonumber
Z_m &=& 1 + \alpha_s(-\frac{2}{\pi}\log(\mu a) + C + b(am_0)^2 + \ldots) + \ldots.
\label{eq:zm}
\end{eqnarray}
From these two equations it is clear that
\begin{equation}
\frac{\overline{m}_c(\mu)}{\overline{m}_s(\mu)} = \left. \frac{am_{0c}}{am_{0s}} \right|_{{\mathrm {phys}}},
\end{equation}
where ${\mathrm{phys}}$ denotes extrapolation to the continuum limit and 
physical sea quark mass limit.

On each ensemble the ratios we have for $am_{0c}/am_{0s}$ then differ 
from the physical value because of three effects: 
mistuning from the correct physical meson 
mass; 
finite $a$ effects that need to be extrapolated away 
and effects because the sea light quark masses 
are not correct. 
We incorporate these into our fitting function:
\begin{align}\label{eq:fit}
	\left.\frac{m_{0c}}{m_{0s}}\right|_\mathrm{lat} &=
	\left.\frac{m_{0c}}{m_{0s}}\right|_\mathrm{phys} \times
	\left(
	1 + d_\mathrm{sea}\frac{\delta m^\mathrm{sea}_\mathrm{tot}}{m_s}
	\right) \\ \nonumber
	&\times
	\left(
	1 + \sum_{i,j,k,l} c_{ijkl} \,\delta_c^i\,\delta_s^j
	\left(\frac{am_{\eta_c}}{2}\right)^{2k}
	(am_{\eta_s})^{2l}
	\right).
\end{align}	
\begin{equation}
\delta_c = \frac{m_{\eta_c, MC} - m_{\eta_c, \mathrm{phys}}}{m_{\eta_c,\mathrm{phys}}};\, 
\delta_s = \frac{m^2_{\eta_s, MC} - m^2_{\eta_s, \mathrm{phys}}}{m^2_{\eta_s,\mathrm{phys}}}
\end{equation}
are the measures of mistuning, where $MC$ denotes lattice values 
converted to physical units. 
The last bracket fits the finite lattice spacing effects as a power 
series in even powers of $a$. These can either have a 
scale set by $m_c$ (for which we use $am_{\eta_c}/2$) or by $\Lambda_{QCD}$ 
(for which we use $am_{\eta_s}$). $i,j,k,l$ all start from zero and 
are varied in the ranges: $i, j \le 3$, $k \le 6$, $l \le 2$ with $i+j+k+l \le 6$. 
Doubling any of the upper limits has negligible effect on the final result. 
The prior on $c_{ijkl}$ is set to 0(1). 
$\delta m^{\mathrm{sea}}_{\mathrm{tot}}$ is the total difference between 
the sea-quark masses used in the simulation and the correct value for 
$2m_l+m_s$~\cite{newr1}. This has a tiny effect
and we simply use a linear term (adding 
higher orders has negligible effect). 
The prior for $d_{\mathrm{sea}}$ is 0.0(1).   
Figure~\ref{fig:mcmsplot} shows the results of the fit, giving
$m_c/m_s$ in the continuum limit as 11.85(16) ($\chi^2/{\mathrm{dof}}$ = 0.42).  
The error budget is given in 
Table~\ref{tab:errormcms}. 

\begin{table}
\begin{tabular}{lll}
\hline
& $m_c/m_s$ & $m_s/m_l$ \\
\hline
      overall $r_1$ uncertainty &  0.4\% &  0.1\% \\
          $r_1/a$ uncertainties &  0.2 & - \\
         continuum $M_{\eta_c}$ &  0.2 &  - \\
         continuum $M_{\eta_s}$ &  1.1 &  - \\
          Finite volume & - & 0.3 \\
$a^2$ extrapolation, $m_q$ interpolns &  0.4 & 0.8 \\
   sea-quark mass extrapolation &  0.0 &  0.2 \\
             statistical errors &  0.3 &  0.4 \\
\hline
                Total  &  1.3\%  &  1.0\%\\

\hline
\hline
\end{tabular}
\caption{ Error budgets for $m_c/m_s$ and $m_s/m_l$. 
}
\label{tab:errormcms}
\end{table}

$m_s/m_l$ is known to 1\% from lattice QCD as a byproduct of 
standard chiral 
extrapolations of $m_{\pi}^2$ and $m_K^2$ to the 
physical point~\cite{milcm}. MILC quote 27.2(3) using asqtad quarks~\cite{milcreview}. 
Our HISQ analysis in~\cite{fds} gave a result in agreement at 27.8(3),  
using a Bayesian fit 
to a function including terms from chiral perturbation theory
up to third order in $m_l$ and
allowing for discretisation errors up to and including $a^4$ 
and for mixed terms (i.e $m_l$-dependent discretisation 
errors). 
A full error budget is given in 
Table~\ref{tab:errormcms}; the data are given in~\cite{newr1}.

{\it Conclusions}. -- Our $m_c/m_s$ can be used with any value for $m_c$ to give $m_s$. 
The best existing result~\cite{mcjj} (converted from 
$n_f$=4 to 3) is
$\overline{m}^{(3)}_c(2.0{\rm GeV})$ = 1.095(11) GeV or
$\overline{m}^{(3)}_c(3.0{\rm GeV})$ = 0.990(10) GeV. 
Dividing by 11.85(16) gives
$\overline{m}^{(3)}_s(2.0{\rm GeV})$ = 92.4(1.5) MeV and 
$\overline{m}^{(3)}_s(3.0{\rm GeV})$ = 83.5(1.4) MeV. 

\begin{figure}[h]
\includegraphics[width=6.5cm]{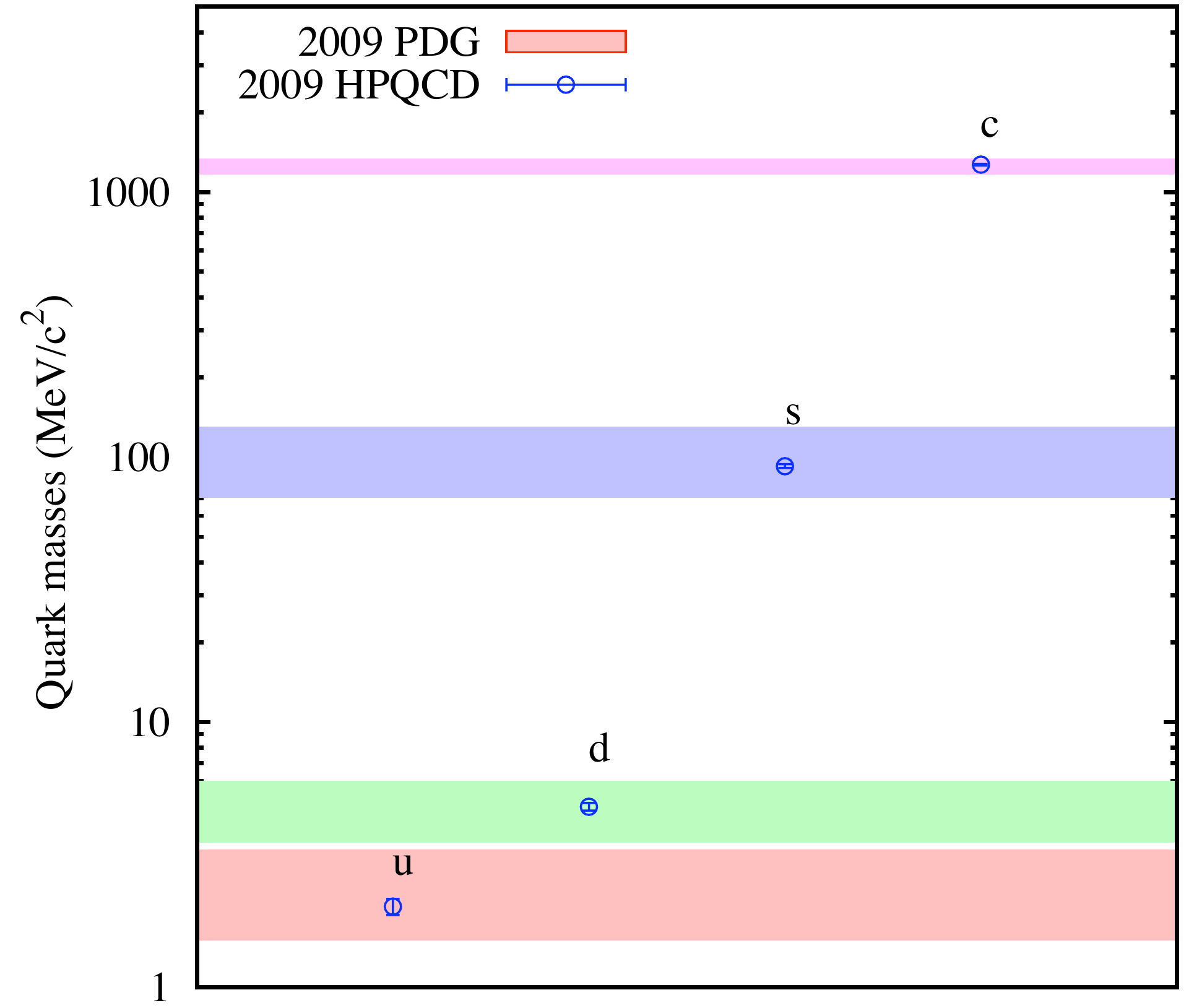}
\caption{ Our results for the 4 lightest quark masses 
compared to the current PDG evaluations (shaded bands)~\cite{pdg09}. 
Each mass is quoted in 
the $\overline{MS}$ scheme at 
its conventional scale: 2 GeV for $u$, $d$, $s$ ($n_f=3$); $m_c$ for 
$c$ ($n_f=4$).}
\label{fig:qmass}
\end{figure}

Using the MILC values for $m_s/m_l$ and $m_u/m_d$ (0.42(4)~\cite{milcreview}) 
we can then obtain: 
$\overline{m}^{(3)}_l(2.0{\rm GeV})$ = 3.40(7) MeV and 
$\overline{m}^{(3)}_l(3.0{\rm GeV})$ = 3.07(6) MeV; 
$\overline{m}^{(3)}_u(2.0{\rm GeV})$ = 2.01(14) MeV 
and $\overline{m}^{(3)}_d(2.0{\rm GeV})$ = 4.79(16) MeV. 
The values for all four quark masses are plotted 
in Figure~\ref{fig:qmass} in comparison to the 
current evaluations from the Particle Data Tables~\cite{pdg09}. 

Thus our high accuracy on $m_c/m_s$ allows us 
to leverage 2\% accurate values for $m_s$ and $m_l$ that 
are completely nonperturbative in lattice QCD, for the first time.
Our $m_s$ mass is higher, by 
around $1 \sigma$, than our previous value 
of $\overline{m}_s(2 {\rm GeV})$ = 87(6) MeV which used
2-loop lattice QCD perturbation theory~\cite{ourstrange}.
Then the error was dominated by unknown $\alpha_s^3$
terms. Our new result, which does not have this limitation, 
has an error almost five times smaller. 
Our new error is almost an order of magnitude smaller 
than other lattice QCD results from full QCD~\cite{rbc, jlqcd}. These 
use direct methods of converting the lattice mass 
to the $\overline{MS}$ mass, and have 10\% errors.  

{\bf{Acknowledgements}} We are grateful to MILC for 
configurations and to H. Leutwyler for useful discussions. 
Computing was done at the Ohio Supercomputer Centre, USQCD's Fermilab cluster and at the Argonne 
Leadership Computing Facility supported by DOE-AC02-06CH11357. 
We used chroma for some analysis. 
We acknowledge support by 
the Leverhulme Trust, the Royal Society, STFC, SUPA, MICINN, NSF and DoE.


\begin{thebibliography}{99}
\bibitem{pdg09} Particle Data Group, http://pdg.lbl.gov/. 
\bibitem{pert} Y. Kiyo {\it et al}, Nucl. Phys. B{\bf 823}:269 (2009) [arXiv:0907.2120]; A.Hoang {\it et al}, Nucl. Phys. B{\bf 813}:349 (2009) [arXiv:0807.4173]. 
\bibitem{kuhn09} J. H. K\"{u}hn, M. Steinhauser and C. Sturm, Nucl. Phys. B{\bf 778}, 192 (2007) [arXiv:hep-ph/0702103]; K. G. Chetyrkin {\it et al} [arXiv:0907.2110]. 
\bibitem{mcjj} I. Allison {\it et al}, HPQCD + K. G. Chetyrkin, J. H. K\"{u}hn, M. Steinhauser and C. Sturm, Phys. Rev. D{\bf 78}:054513 (2008) [arXiv:0805.2999]. 
\bibitem{mylat08} C. T. H. Davies {\it et al}, HPQCD, PoS(LATTICE 2008)118 [arXiv:0810.3548]. 
\bibitem{mchighbeta} I. Allison {\it et al}, HPQCD, PoS(LATTICE 2008)225 [arXiv:0810.0285].  
\bibitem{gasserleutwyler} J, Gasser and H. Leutwyler, Phys. Rept. {\bf 87} (1982) 77. 
\bibitem{ratio} C. T. H. Davies {\it et al}, HPQCD/Fermilab/MILC, Phys. Rev. 
Lett. {\bf 92}:022001 (2004) [arXiv:hep-lat/0304004].
\bibitem{lp07} E. Gregory {\it et al}, HPQCD, Phys. Rev. Lett. {\bf 104}:022001 (2010) [arXiv:0909.4462].
\bibitem{ourstrange} Q. Mason {\it et al}, HPQCD, Phys. Rev. D{\bf 73}:114501 (2006) [arXiv:hep-ph/0511160]. 
\bibitem{hisq} E. Follana {\it et al}, HPQCD, Phys. Rev. D{\bf 75}:054502 (2007) [arXiv:hep-lat/0610092]. 
\bibitem{fds} E. Follana {\it et al}, HPQCD, Phys. Rev. Lett.{\bf 100}:062002 (2008) [arXiv:0706.1726].
\bibitem{sharpe} S. Sharpe, PoSLAT2006:022 [arXiv:hep-lat/0610094].
\bibitem{milcreview} A. Bazavov {\it et al}, arXiv:0903.3598.
\bibitem{kronfeld} A. S. Kronfeld, PoSLAT2007:016 [arXiv:0711.0699].
\bibitem{golterman} M. Golterman, PoSCONF8:014 [arXiv:0812.3110].
\bibitem{bayesfits} G. P. Lepage {\it et al}, Nucl. Phys. B (Proc. Suppl. {\bf 106}), 12 (2002) [arXiv:hep-lat/0110175].
\bibitem{newr1} C. T. H. Davies {\it et al}, HPQCD, arXiv:0910.1229, Phys. Rev.D (in press).
\bibitem{massnote} Note that~\cite{mcjj} determined $m_c$ using an earlier 
and less accurate value of $r_1$ (0.321(5) fm). We have checked that using the 
new value does not change $m_c$ significantly. 
\bibitem{mctree} 
The mass parameter $am_{0c}$ in equation~\ref{eq:zm} 
should really be the perturbative 
pole mass at tree level~\cite{hisq}, which differs from $am_{0c}$ by $\approx 0.04(am_{0c})^4$, 
but gives the same extrapolated $m_c/m_s$ ratio. 
\bibitem{milcm} C. Aubin {\it et al}, MILC, Phys. Rev. D{70}:114501 (2004) [arXiv:hep-lat/0407028].
\bibitem{rbc} C. Allton {\it et al}, RBC-UKQCD, Phys. Rev. D{\bf 78}:114501 (2008) [arXiv:0804.0473].
\bibitem{jlqcd} T. Ishikawa {\it et al}, CP-PACS/JLQCD, Phys. Rev. D{\bf 78}:011502R (2008) [arXiv: 0704.1937].
\end{thebibliography}
\end{document}